\documentclass[]{article}
\usepackage{dcolumn}
\usepackage{natbib}
\usepackage{amsmath}
\usepackage{amssymb}
\usepackage{listings}
\usepackage[T1]{fontenc}
\usepackage[utf8]{inputenc}
\usepackage{graphicx}
\usepackage{epstopdf}
\usepackage{color}
\usepackage{float}
\usepackage{tikz}
\usepackage{tcolorbox}
\usepackage{multirow}
\usepackage{hhline}
\usepackage{xcolor}
\usepackage{colortbl}
\usepackage{subcaption}
\usepackage{booktabs}
\usepackage{lscape}
\usepackage{subcaption}
\usepackage{pgfplots}
\usepackage[utf8]{inputenc}
\usepackage[letterpaper,pdftex]{geometry}
\usepackage{longtable}
\usetikzlibrary{arrows,automata,fit,scopes,calc,matrix,positioning,decorations.pathmorphing,decorations.pathreplacing}
\pgfplotsset{width=9cm,compat=1.14}
\usepgfplotslibrary{statistics}
\usepackage[all]{xy}
\usepackage[affil-it]{authblk}
\usepackage[doublespacing]{setspace}
\usepackage{rotating}

\usepackage{hyperref}	
\hypersetup{colorlinks,linkcolor={blue},citecolor={blue!50!black},urlcolor={blue}}
\usepackage{geometry}

\title{The Impact of Birth Order on Behavior in Contact Team Sports: the Evidence of Rugby Teams in Argentina}
\author[1]{Fernando Delbianco \footnote{fernando.delbianco@uns.edu.ar}} 
\author[1]{Federico Fioravanti \footnote{federico.fioravanti9@gmail.com}}
\author[1]{Fernando Tohm\'e \footnote{ftohme@criba.edu.ar}}

\affil[1]{INMABB, Universidad Nacional del Sur, Bah\'{\i}a Blanca, Argentina}
\date{}
\begin{document}

\maketitle

\begin{abstract}
Several studies have shown that birth order and the sex of siblings may have an influence on individual behavioral traits. In particular, it has been found that second brothers (of older male siblings) tend to have more disciplinary problems. If this is the case, this should also be shown in contact sports. To assess this hypothesis we use a data set from the South Rugby Union (URS) from Bahía Blanca, Argentina, and information obtained by surveying more than four hundred players of that league. We find a statistically significant positive relation between being a second-born male rugby player with an older male brother and the number of yellow cards received. 
\end{abstract}

\section{Introduction}

The understanding of human behavior is fundamental in the analysis of intentional interactions, particularly in regulated environments, ranging from societies under the rule of law to institutions operating under implicit rulings. Traditionally, this kind of study was carried out under the assumption that individuals are {\em rational}, i.e. that they behave consistently with their preferences. But since Herbert Simon introduced the concept of {\em bounded rationality}, the idea that rational behavior is modulated and even reshaped by factors independent of the conscious control of individuals has gained widespread acceptance (Simon 2000).

The tools of Big Data have been instrumental in the detection of those ``behavior-modifying'' factors, hidden but present in large reams of data. On the other hand, once detected, it is of great interest to find how they are manifested in specific contexts. In the case of organizations such evidence can be relevant for redesigning and improving them.

In this paper we take up from a recent study, based on the analysis of large databases on the life records of individuals in Florida and Denmark, that shows that the position in the birth order has an impact on delinquency outcomes (Breining et al. 2020). Other investigations have extended this result to other aspects of human life (Black et al. 2018) (Ginja et al. 2020) (Esposito et al. 2020). Birth order seems, thus, to be an explanatory variable for a large variety of capacities (and shortcomings) at play in individual behaviors.

The evidence indicates that males with older brothers tend to exhibit more unruly and aggressive conducts. We speculate that such behavior should be evidenced in the specific context of contact sports. Players that are second brothers should exhibit, in average, a large tendency to incur offenses. Those players, if this behavioral trait is assumed, should then exhibit a larger record of yellow and red cards.

To evaluate this hypothesis we use data on more than 400 male players of the different teams that play in the South Rugby Union (Uni\'on de Rugby del Sur - URS), Argentina. We use information drawn from the records of the 2019 season of the local and regional championships as well as from a questionnaire on the position of the players in the birth order of their families. 

Based on this evidence we show that, indeed, it is the case that second-born male players with male older brother, tend to commit a statistically significant larger number of offenses. 

The plan of this article is as follows. In Section 2 we briefly discuss the literature on the influence of birth order on behavior. In Section 3 we describe the essential features of Rugby that are relevant for this study. In Section 4 we discuss the database on which we run our analysis and the way in which the information was collected. Section 5 presents the results. Finally, in Section 6 we conclude.

\section{Birth order and behavior}

The relevance of birth order as a psychological variable has been discussed for decades, but no definite impact on behavioral traits were found at the early stage of those investigations. So for instance, while firstborns were shown to exhibit higher levels of conformity and need for achievement (Forer, 1977), other relevant variables that could also explain this were not explored. Reviews by Ernst and Angst (1983) and Dunn and Plomin (1990) found little reliable evidence of this relations: ``[birth] order does not appear to be a very strong influence in molding personality in a definable way'' (Ernst and Angst  1983).

In a very influential book, Sulloway (1996) acknowledged that even if birth-order effects could be subtle, they might be detected in very large samples. While his results were controversial, this opened the door for further studies on issues like the influence of birth order on early smoking (Bard and Rodgers 2006) or on intelligence (Darmian and Roberts 2015). A long term study showed that birth order has an impact on differences in health and educational attainment of older and younger siblings (Barclay and Kolk 2018).

A common finding in these recent studies is a slight but statistically significant difference between second-born boys that tend to exhibit worse results than older brothers. Breining et al. (2020) carried out a large study, using lifelong data from people of diverse cultural and ethnic backgrounds in Denmark and the state of Florida, detecting that younger brothers tend to exhibit a higher level of delinquent behavior. One of the main explanations is the time devoted by parents to their first born children in comparison with the effort spent in their younger siblings. Interestingly, this is not as marked when either the older or the youngest child is a girl.

This leads us to the immediate conclusion that this effect should manifest in contact sports, in which there are numerous opportunities for wrongdoing. While there are different sports in which this effect could be at play, in some of them (soccer, for instance), the impact of birth order may be confounded by cultural and socio-economic differences, that may also lead to aggressive behavior. To avoid those confusions we decided to study how this effect plays out in a sport in which the background of the players is more culturally homogeneous and relatively prosperous. For this reason we have chosen rugby, a game played in Argentina by middle and upper-class people of European-influenced culture (Bautista Branz 2016).

\section{The game of Rugby}

Rugby is a contact sport, that has grown consistently through the years to be currently played by more than 8.5 million people worldwide, aged from six to 60+. The variety of skills and physical conditions that Rugby requires results in a diversity of body shapes, sizes and abilities. The game is played by two teams of 15 players each one\footnote{There is a faster version of the game in which the squads are of seven players but play in the same area, during two halves of seven minutes.}. They play in two half times of 40 minutes in a rectangular shaped field. The objective of the game is to score more points than the opposing team. There are five methods of scoring: with a try, which awards 5 points; with a conversion (2 points); a penalty try (7 points); a penalty goal or a dropped goal (3 points each).\footnote{A try is scored when an attacking player grounds the ball in the opponents in-goal; a conversion is a kick to the posts after a try; a penalty try is awarded after an infringement of the defending team prevents a probable try from being scored; a penalty goal is a kick to the posts after and infringement and a dropped goal is a kick to the posts during play.} Players are classified in two big groups, the forwards and the backs. Traditionally, the forwards are the players that seek to get the possession of the ball while the backs are the ones that score tries. Forwards are usually heavy players, with strength to dispute the ball, while backs are light players, with speed to score tries.

The key features of a good rugby match are the fair contest for the ball and the continuity of the game. Every action that attempts against these two key features, the security of the players or the values of the game (integrity, discipline and respect) is punished. When an offense committed by a player is serious, a yellow or red card is shown to him. Every foul play like punching, kicking, spitting, etc, is usually penalized with a yellow or red card (penalizing actions against security and values). Another offense that is penalized with a card, is when a player or a team infringes the rules many times or when their infringement prevents the other team to score a try (penalizing actions against fair contest and continuity). A typical situation where many offenses occur is the breakdown\footnote{The period of time after a tackle and during the ensuing ruck (a contest situation where two or more stand players dispute the ball).}. On a rugby game can happen from $100$ to $150$ breakdown situations\footnote{{\tt https://www.world.rugby}.} and many players are involved in them, specially the forwards. That is why many of the offenses of a rugby game happen in this context.

Therefore, a yellow or a red card indicates, in increasing degree, that a player has failed to act according to the values of the game and can be seen as a sign of unruly behavior (Romand and Pantal\'eon 2007).

\section{Data}
The data for this research has been collected from the records of the South Rugby Union (URS) of Argentina and from the answers to a survey conducted by the URS on the family structure and education of the players. We will discuss these two sources and the information extracted from them.

The URS provided us the following data about the players:
\begin{itemize}
\item Full name
\item Club
\item Date of birth
\item Height and weight
\item Position on field
\item Yellow and red cards record
\end{itemize}

This information was later matched to the responses given by each player to a survey\footnote{The survey questionnaire can be found in Appendix 1.}. The questions asked to them were the following:

\begin{itemize}
\item Years playing rugby
\item Educational attainment
\item Birth order
\item Siblings gender
\end{itemize} 

The survey was sent to the players through the coaches of the teams of the URS. The devolution rate was of around $60\%$ ($415$ completed surveys out of $698$ senior players).\footnote{We only consider the complete surveys, i.e., the ones with all the questions answered.} The respondents were all male amateur rugby players of ages $18+$ playing in the local and regional tournaments recorded in the URS database. These players belonged to $21$ different clubs from the South of the province of Buenos Aires, the East of La Pampa and the North of R\'{\i}o Negro. 

We decided to disregard the answers of younger players, because in the junior competitions in which they participate the refereeing is not strict. In these matches it happens frequently that even if a player may deserve a yellow or red card, he may not be shown one.

\section{Descriptive statistics}

In Table~\ref{tab:summ} we can see the main statistical description obtained by combining the URS database with the survey answers.\footnote{We used {\bf Gretl} (\texttt{http://gretl.sourceforge.net/}) to perform the statistical and econometric analyses in this article.}  We use two indicator variables that contribute to answering our research question, \emph{Code 1} and \emph{Code 2}. The first one is $1$ if the player is a second male child with an older male brother and $0$ otherwise. \emph{Code2} is also a dummy variable, which indicates whether or not a player is a second male child but with an older sister. 

We decided to consider only yellow cards because very few red cards are recorded in the database, as seen in the $0.0169$ mean value of the {\emph red cards} variable in Table~\ref{tab:summ}. The database is further completed with contextual information, such as club membership, education (as incomplete secondary education, complete secondary school, incomplete college and complete tertiary education), the position and category corresponding to each player. This extra information is used to define different indicator variables for control purposes in our regressions. Among them we include for each player the position in the team (forward, back or wing), to which club he belongs and the category or level at which his team plays.

\begin{table}[!htbp] 
	\centering 
	\caption{Summary statistics }
	\label{tab:summ}
	\begin{tabular}{l c c c c c c}
		Variable & Mean & Median & Std. Dev & Min
		& Max & Observations \\[1ex]
		\hline
		\hline
		\emph{Age} & 25.9 & 25.0 & 6.47 & 18.0 & 50.0 & 415  \\
		{\em Yellow Cards} & 0.667 & 0.000 & 1.06 & 0.000 & 8.00 & 415 \\
		{\em Red Cards} & 0.0169 & 0.000 & 0.129 & 0.000 & 1.00 & 415 \\
		{\em Weight} & 90.0 & 88.0 & 17.1 & 51.0 & 145.0 & 310 \\
		{\em Height} & 1.77 & 1.78 & 0.0675 & 1.55 & 1.98 & 310 \\
		\emph{Code 1} & 0.186 & 0.000 & 0.389 & 0.000 & 1.00 & 415 \\
		\emph{Code 2} & 0.116 & 0.000 & 0.320 & 0.000 & 1.00 & 415 \\
		\hline
	\end{tabular}
\end{table}

We focus on the variables that, as will be shown in the next section, are statistically relevant. That is, we will here give a description of the distribution of the number of yellow cards received, the dummy variables corresponding to birth order and the weight of players, as indicated in Table~\ref{tab:summ}. 

In Figure~\ref{fig:hist} we can see the distribution of the total number of yellow cards, i.e. the number of players with $0, 1, \ldots$ up to $8$ cards. The left panel corresponds to the players for which \emph{Code 1}$=$\emph{Code 2}$=0$, while the middle one to those with \emph{Code 1}$=1$ and the right one to those with \emph{Code 2}$=1$.

We can see in the middle panel that the proportion of one and two yellow cards with respect to $0$ cards, is much larger than in the cases of the left and the right panel. Furthermore, these two panels show similar distributions.
 
An interesting collateral result is the relation between \emph{Code 1} and \emph{Code 2}  and the weight variable, as we can see in Figure~\ref{fig:boxplot}. The middle panel corresponding to \emph{Code 1}$=1$, shows a higher median, more skewed to the right, in comparison to the bottom (\emph{Code 1}$=$\emph{Code 2}$=0$) and top (\emph{Code 2}$=1$) boxplots.  

\begin{figure}[hbt!]
	\centering
		\caption{Number of yellow cards (x-axis) and number of cases in the database (y-axis).}
	\begin{subfigure}[t]{0.3\textwidth}
		\begin{tikzpicture}[xscale=0.7]
		\begin{axis}[ybar interval, x=0.444cm, y=0.0277cm, xmax=9, xmin=0, ymax=180, ymin=0, minor y tick num = 3]
		\addplot coordinates { (0, 176) (1, 75) (2, 23) (3, 9) (4, 4) (5, 1) (6, 1) (7, 0) (8, 1) (9,0)};
		\end{axis}
		\end{tikzpicture}
		\caption{}
	\end{subfigure}%
	~ 
	\begin{subfigure}[t]{0.3\textwidth}
		\begin{tikzpicture}[xscale=0.7]
		\begin{axis}[ybar interval, x=0.667cm, y=0.125cm, xmax=6, xmin=0, ymax=40, ymin=0, minor y tick num = 3]
		\addplot coordinates { (0, 32) (1, 27) (2, 10) (3, 3) (4, 4) (5, 1) (6,0)};
		\end{axis}
		\end{tikzpicture}
		\caption{}
	\end{subfigure}
	~ 
	\begin{subfigure}[t]{0.3\textwidth}
		\begin{tikzpicture}[xscale=0.7]
		\begin{axis}[ybar interval, x=0.8cm, y=0.125cm, ymax=40,ymin=0, xmax=5, xmin=0, minor y tick num = 3]
		\addplot coordinates { (0, 37) (1, 7) (2, 2) (3, 2) (4, 0) (5,0)};
		\end{axis}
		\end{tikzpicture}
		\caption{}
	\end{subfigure}	

	\caption*{(a): $Code 1= Code 2=0$; (b): $Code 1=1$; (c): $Code 2=0$}		
	\label{fig:hist}
\end{figure}

In Figure~\ref{fig:boxplot} we can see that the median weight corresponding to {\em Code 1} is larger (93 kilograms) than that of the other groups, with a wider inter-quartile range, indicating a larger dispersion.  

\begin{figure}[htbt!]
	\centering
	\caption{Box Plot of {\em Weight}.}
\begin{tikzpicture}
\centering
\begin{axis}
\addplot+ [
boxplot prepared={
	lower whisker=55.75, lower quartile=78,
	median=88,
	upper quartile=100.25, upper whisker=122.25,
},
] table [row sep=\\,y index=0] { 51\\ 130\\ 132\\ 133\\ 145\\ };
\addplot+ [
boxplot prepared={
	lower whisker=52, lower quartile=76,
	median=93,
	upper quartile=100, upper whisker=124,
},
]
coordinates { (0,125) (0,130)};
\addplot+ [
boxplot prepared={
	lower whisker=54.5, lower quartile=76,
	median=85,
	upper quartile=97.5, upper whisker=119,
},
] coordinates {(0,125) (0,138)};

\end{axis}
\end{tikzpicture}
\caption*{Note: top: \emph{Code 2}$=1$ ; middle: \emph{Code 1}$=1$; bottom: \emph{Code 1}$=$\emph{Code 2}$=0$.}
\label{fig:boxplot}
\end{figure}

\begin{figure}[hbt!]
	\centering
	\caption{Yellow cards and Weight}		
	\begin{subfigure}[t]{0.47\textwidth}
		\centering
		\scriptsize
		\includegraphics[width=1\textwidth]{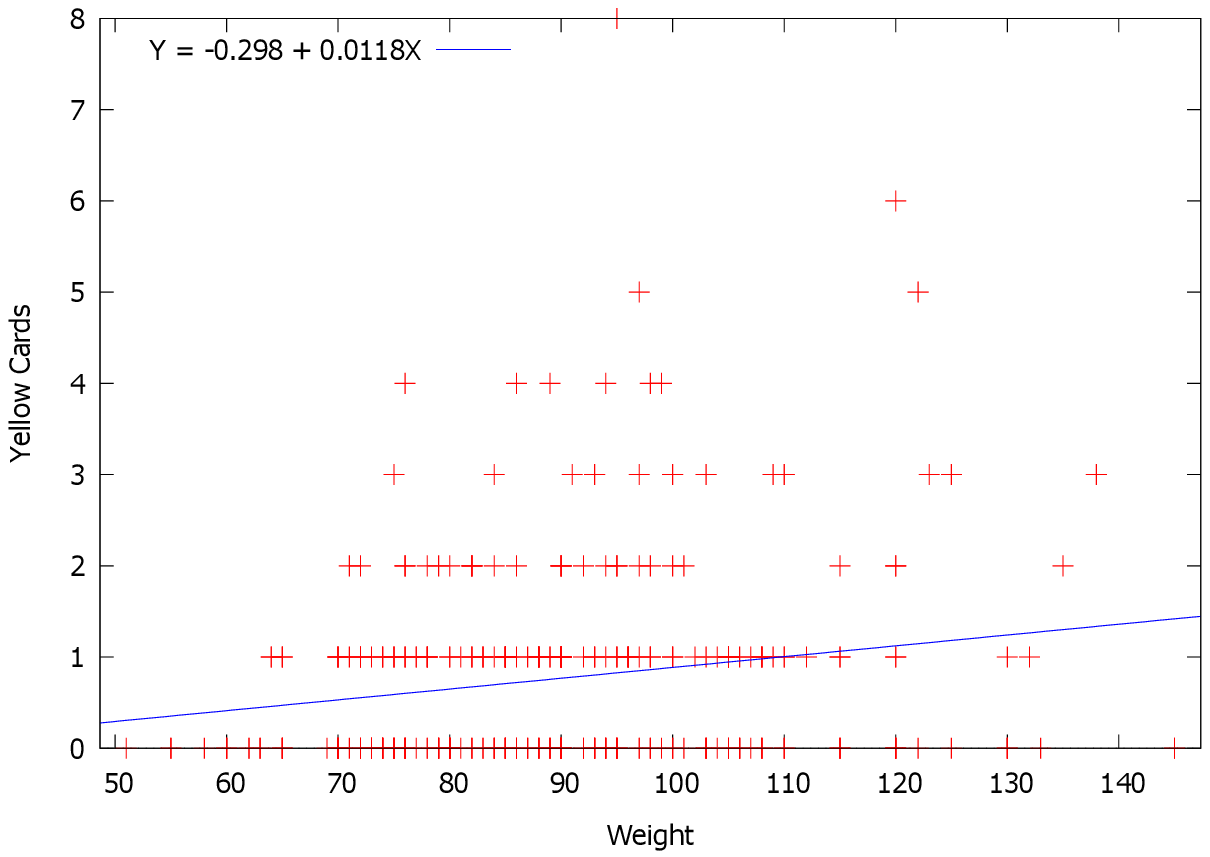}
		\caption{Linear relation. \label{fig:lineal}}
	\end{subfigure}%
	~ 
	\begin{subfigure}[t]{0.47\textwidth}
		\centering
		\scriptsize
		\includegraphics[width=1\textwidth]{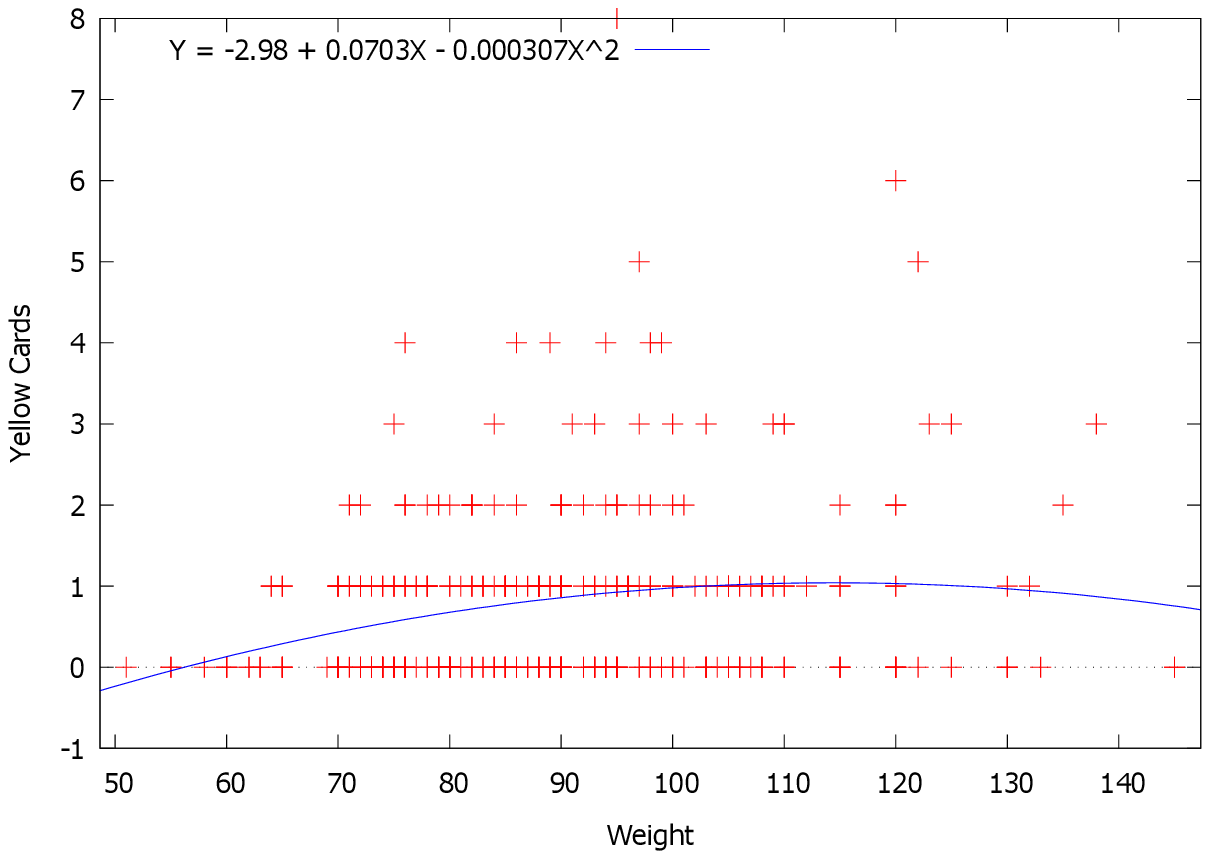}
		\caption{Quadratic relation. \label{fig:cuadratica}}
	\end{subfigure}
\end{figure}

We can see the relation between the number of {\em Yellow Cards} received by a player and his (declared) {\em Weight} in Figures~\ref{fig:lineal} and \ref{fig:cuadratica}, indicating a positive relation between the weight of players and the number of yellow cards they receive, although at the largest weights (very few cases) this relation is lost. 

Finally, we test for the variables of interest the differences between means, as reported in Table~\ref{tab:mean}. Firstly, there exist a significant difference between the mean of {\em Yellow Cards} for {\em Code 1}$=1$ (an average of $1$ card), and the $0.59$ mean of the rest of the sample. The second test shows, despite the evidence of the boxplot for medians, that when we test the difference of means of {\em Weight} under {\em Code 1}$=1$, against the rest of the sample, there are no significant differences. Finally, in the third test, we compare the mean weight of the players that did not get yellow cards with that of those who got at least $1$ yellow card. The statistically significant difference is of almost $4$ kilograms.

\begin{table}[!htbp] 
	\centering 
	\caption{Mean tests}
	\label{tab:mean}
	\begin{tabular}{r r l r l r l}
		\hline
		 & \multicolumn{2}{c}{Test 1} & \multicolumn{2}{c}{Test 2}
		& \multicolumn{2}{c}{Test 3} \\[1ex]
	
		Mean: & \multicolumn{2}{c}{{\em Yellow Cards}} & \multicolumn{2}{c}{{\em Weight}}
		& \multicolumn{2}{c}{{\em Weight}} \\[1ex]
		
		\multirow{ 2}{*}{Condition:} & \multicolumn{2}{c}{{\em Code 1}$=1$ vs.} & \multicolumn{2}{c}{{\em Code 1}$=1$ vs. }
		& \multicolumn{2}{c}{{\em Yellow Cards}$=0$ vs. } \\[1ex]
		 & \multicolumn{2}{c}{{\em Code 1}$=0$} & \multicolumn{2}{c}{ {\em Code 1}$=0$}
		& \multicolumn{2}{c}{ {\em Yellow Cards}$>0$} \\[1ex]
		\hline
		\hline
		\multirow{ 3}{*}{Mean (group 1)} & n & $77$ & n & $53$ & n & $163$  \\
			 & mean & $1.00$ & mean & $89.1509$ & mean & $88.1472$ \\
			 & sd & $0.13$ & sd & $2.14435$ & sd & $1.42932$ \\
		\hline
		\multirow{ 3}{*}{Mean (group 2)} & n & $338$ & n & $257$ & n & $147$ \\
		     & mean & $0.5917$ & mean & $90.1673$ & mean & $92.0408$ \\
		     & sd & $0.0555$ & sd & $1.08896$ & sd & $1.28824$ \\
		\hline
		\emph{Two tails p-value} &  & $0.002271$ &  & $0.6949$ &  & $0.04558$ \\
		\emph{One tail p-value} &  & $0.001135$ &  & $0.3474$ &  & $0.02279$ \\
		\hline
	\end{tabular}	
	\caption*{Note: n: number of observations; sd: standard deviation.}
\end{table}

\section{Regression results}

Our main hypothesis, as indicated above, is that the number of yellow cards should be larger for \emph{Code 1} players. That is, second-born males with an older brother will tend to commit more offenses. To evaluate this hypothesis we ran a series of regressions of the form:

\begin{equation}
Yellow \ Cards_i = \beta_0 + \beta_1 \mbox{\emph{Code 1}}_i + \beta_2 \mbox{\emph{Code 2}}_i + \Gamma X_i + \epsilon_i
\end{equation}

\noindent in which our coefficients of interest are $\beta_1$ and $\beta_2$, varying the control variable $X$. Table~\ref{tab:results} summarizes the results of these regressions
 
In the first column we report the results of running the standard OLS with no controls. We can see that the results of \emph{Code 1} and \emph{Code 2} are the expected according to our hypothesis, positive and negative respectively. Not surprisingly, the Breush-Pagan test indicates the presence of heteroskedasticity (Non-constant Variance Score Test with $\chi^2 = 6.965588$, $p = 0.0083092$), while a Cook-distance test reveals the presence of influential observations in the simple regression. Due to these results, we had to refine our analysis. We did so by applying two strategies: by considering White's robust errors and by clustering the variance and covariance matrices. 

In columns $2-5$ we can see the robust error regression results, with the different controls added. The main result in these cases is that \emph{Code 2} loses significance in columns $3$ and $4$, but we find a significant and positive effect of the player's weight on the resulting number of yellow cards (although with a very small coefficient). Another  variable that shows a significant effect is the Incomplete Secondary School dummy. 

The last two columns show the results of running the regressions with a clustered error, based on the variable indicating to which club the player belongs. This is in order to capture the intuition that teammates tend to show similar styles of play (sharing a common coach). The results for \emph{Code 1} and \emph{Code 2} are, again, the expected under the hypothesis. 

Given that our dependent variable, {\em Yellow Cards}, is a counting variable, we take this particular feature into account, modeling the relation both as a Poisson regression and as a negative binomial regression. As pointed out by Orme (2009), Hilbe (2011) and Cameron and Trivedi (2013), it is intuitive to think that in the case of a counting variable, it will follow a Poisson distributions:
\begin{equation}
f\left(y_{i} | \mathbf{x}_{i}\right)=\frac{e^{-\mu_{i}} \mu_{i}^{y_{i}}}{y_{i} !}, \quad y_{i}=0,1,2, \ldots
\end{equation}

\noindent where the mean is expressed as follows:

\begin{equation}
\mathrm{E}\left[y_{i} | \mathbf{x}_{i}\right]=\mu_{i}=\exp \left(\mathbf{x}_{i}^{\prime} \boldsymbol{\beta}\right)
\end{equation}

We adjust the observations to this distribution using a Quasi-Maximum Likelihood estimation with robust standard errors, in the form of:

\begin{equation}
\mathrm{V}_{\mathrm{RS}}\left[\hat{\boldsymbol{\beta}}_{\mathrm{P}}\right]=\left(\sum_{i=1}^{n} \mu_{i} \mathbf{x}_{i} \mathbf{x}_{i}^{\prime}\right)^{-1}\left(\sum_{i=1}^{n}\left(y_{i}-\mu_{i}\right)^{2} \mathbf{x}_{i} \mathbf{x}_{i}^{\prime}\right)\left(\sum_{i=1}^{n} \mu_{i} \mathbf{x}_{i} \mathbf{x}_{i}^{\prime}\right)^{-1}
\end{equation}

\noindent where $(y_{i}-\mu_{i})^{2} = \omega_{i}$, is the weighing factor. The results can be seen in table~\ref{tab:results2}.

An important assumption in the case of Poisson distributions is that the variance equals the mean. A $\chi^2$ test yields that the assumption of equality between the mean and the variance is not valid. Thus, we have to drop the assumption of a single parameter, resorting to a negative binomial regression, by including a parameter of overdispersion, $\alpha$, where $\alpha >0$ (in a Poisson model, $\alpha = 0$). 

Taking into account $\alpha$ the term weighting the variance and covariance matrix is:

\begin{equation}
\omega_{i}=\mu_{i}+\alpha \mu_{i}^{p}
\end{equation}

If $p=1$ the model is called Negative Binomial 1 (NB1), and if $p=2$, Negative Binomial 2 (NB2). We do not have an a priori model of the generation of yellow cards, therefore we just take the results obtained up to this point and choose to work with the NB2 model.\footnote{A non-linear specification of the $\omega_i$s is better at capturing finer distinctions in the data.} In the rightmost columns of Table~\ref{tab:results2} we report the general results with this last model, where:

\begin{equation}
\mathrm{E}\left[Yellow_{i} | \mathbf{x}_{i}\right] = \exp(\beta_0 + \beta_1 \mbox{\emph{Code 1}}_i + \beta_2 \mbox{\emph{Code 2}}_i + \beta_3 Weight_i)
\end{equation}

\begin{figure}[hbt!]
	\centering
	\caption{Yellow cards observed, predicted and residuals.}
	\begin{subfigure}[t]{0.47\textwidth}
		\centering
		\scriptsize
		\includegraphics[width=1\textwidth]{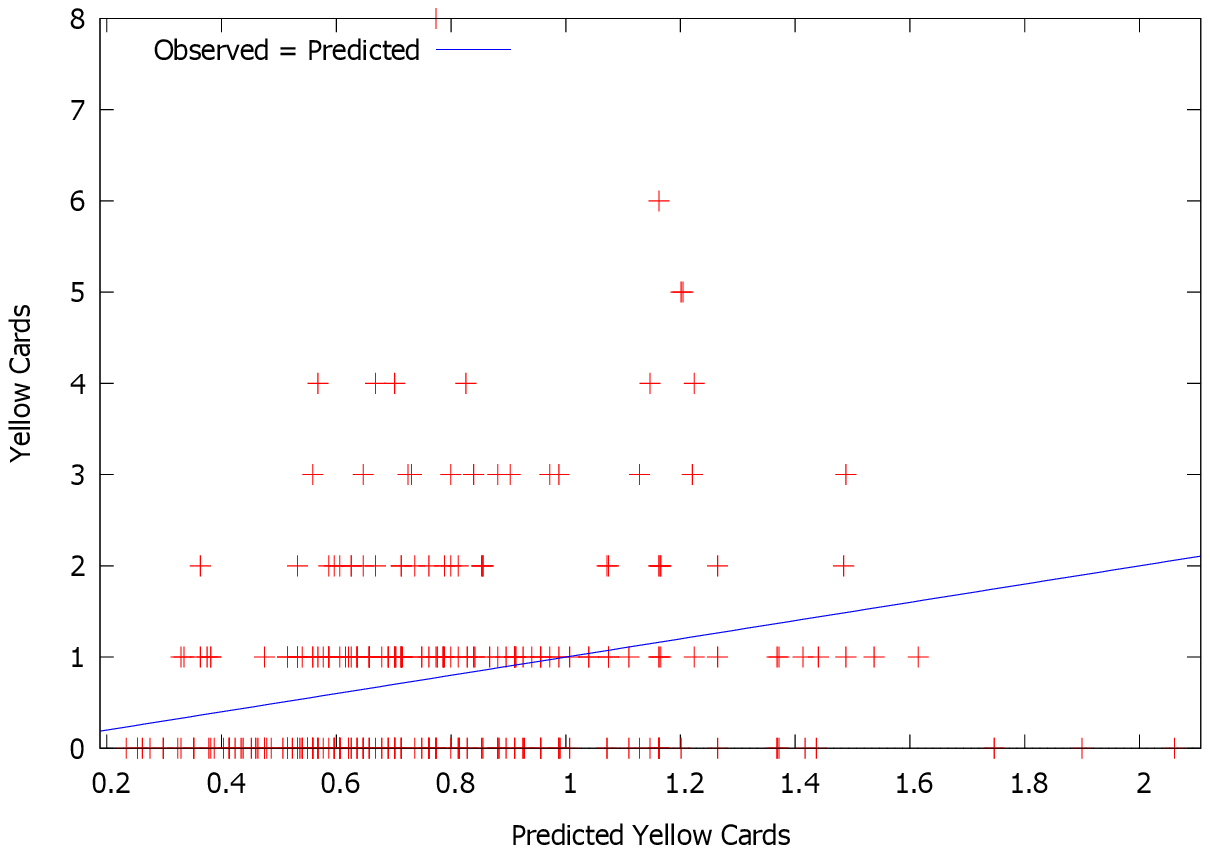}
		\caption{Linear prediction. \label{fig:pred}}
	\end{subfigure}%
	~ 
	\begin{subfigure}[t]{0.47\textwidth}
		\centering
		\scriptsize
		\includegraphics[width=1\textwidth]{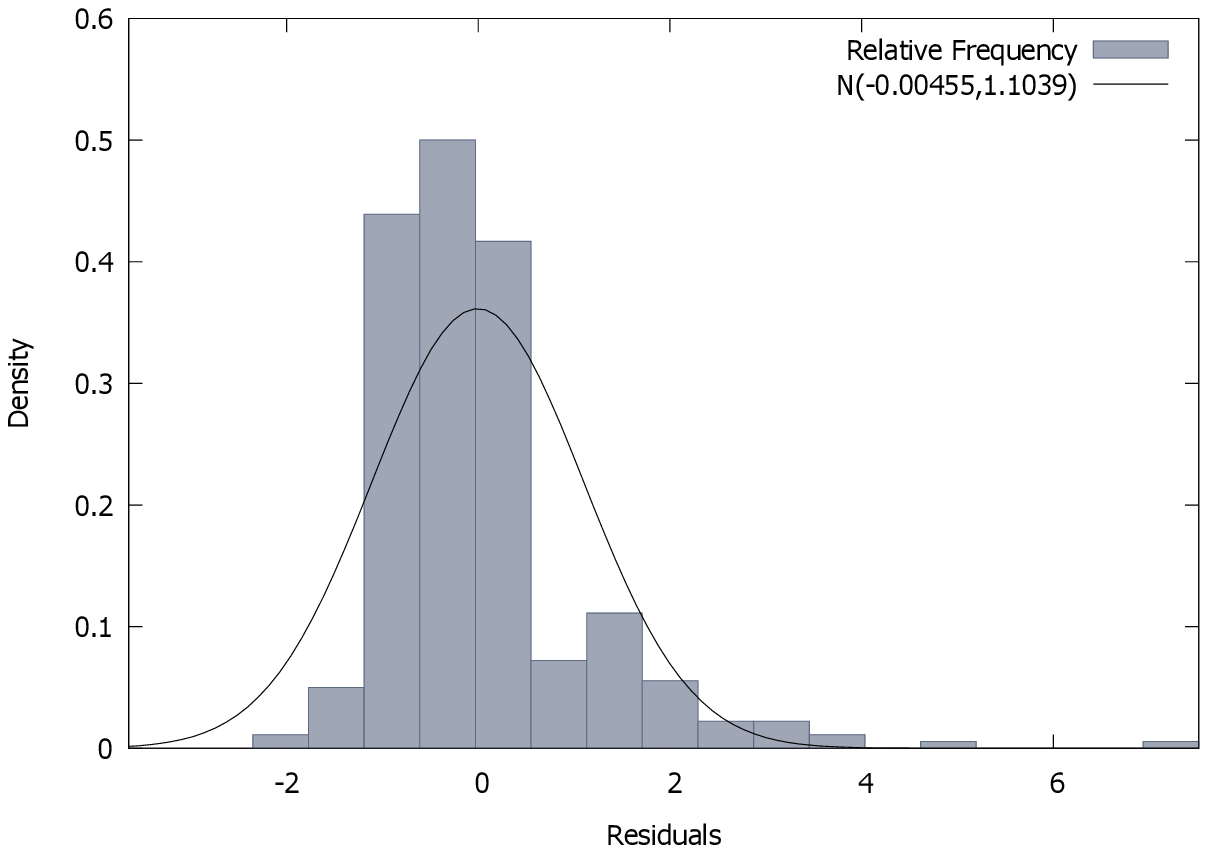}
		\caption{Normality contrast. \label{fig:norm}}
	\end{subfigure}		
\end{figure}

Table~\ref{tab:final_results} and Figures~\ref{fig:pred} and \ref{fig:norm} present the results of this model with our variables of interest. 

\begin{table}[!htbp] \centering 
	\caption{Negative Binomial Regression Results ($p=2$)} 
	\label{tab:final_results} 
	\begin{tabular}{l c c c c}
		& Coefficient & Standard Deviation & $z$ & p-value \\[1ex]
		\hline \hline
		{\em Constant} & $-1.80345$ & $0.439240$ & $-4.106$ & $0.0000$ \\
		{\em Weight} & $0.0162847$ & $0.00468983$ & $3.472$ & $0.0005$ \\
		{\em Code 1} & $0.409718$ & $0.173700$ & $2.359$ & $0.0183$ \\
		{\em Code 2} & $-0.545480$ & $0.295712$ & $-1.845$ & $0.0651$ \\
		$\alpha$ & $0.540983$ & $0.177833$ & $3.042$ & $0.0023$ \\
		\hline
		Observations & $310$ & & & \\
		AIC criteria & $736.9341$ & & & \\
	\end{tabular}
	\caption*{Note: Quasi-Maximum Likelihood Standard Errors.}
\end{table}	

We can see that {\em Code 1}, {\em Code 2}, {\em Weight} and the overdispersion coefficient $\alpha$ are all statistically significant. Figure~\ref{fig:pred} shows that the linear model leaves many actual values of {\em Yellow Cards} far from the predicted ones, resulting in Figure~\ref{fig:norm}, that indicates that most of the residuals are below the normal, being their distribution skewed to the right as a result of the overdispersion. \\
To ensure the robustness of our results, we run this final model with an NB1 specification. The results can be seen in Table \ref{tab:final_results_nb1}. The only difference found is that \emph{Code 2} is not significant. Nonetheless, if we take into account the Akaike information criteria (AIC), the NB2 model, as expected, adjusts better to our data. \\

\begin{table}[!htbp] \centering 
	\caption{Negative Binomial Regression Results ($p=1$)} 
	\label{tab:final_results_nb1} 
	\begin{tabular}{l c c c c}
		& Coefficient & Standard Deviation & $z$ & p-value \\[1ex]
		\hline \hline
		{\em Constant} & $-1.44833$ & $0.409263$ & $-3.539$ & $0.0004$ \\
		{\em Weight} & $0.01233238$ & $0.00431164$ & $2.858$ & $0.0043$ \\
		{\em Code 1} & $0.446600$ & $0.164477$ & $2.715$ & $0.0066$ \\
		{\em Code 2} & $-0.481741$ & $0.318358$ & $-1.513$ & $0.1302$ \\
		$\alpha$ & $0.429182$ & $0.156938$ & $2.735$ & $0.0062$ \\
		\hline
		Observations & $310$ & & & \\
		AIC criteria & $738.7830$ & & & \\
	\end{tabular}
	\caption*{Note: Quasi-Maximum Likelihood Standard Errors.}
\end{table}

Finally, there exists the possibility that {\em Code 1} may be correlated with {\em Weight}, according to the findings of Meller et al. (2018), who find that birth order has also an impact on the health of second brothers. We run thus Hausman test of endogeneity , instrumenting {\em Weight} with {\em Height}. The latter variable is chosen as instrument since its correlation coefficient with {\em Weight} is $0.44565732$, which allows to reject the null hypothesis of no correlation ($t(308) = 8.73684$, with a two tails $p$-value of $0.0000$). Then, Hausman's test yields $\chi^2=0.02$, and thus the null hypothesis of exogeneity cannot be rejected, validating the results reported in table~\ref{tab:final_results}.

\section{Conclusions}

In this work we established a significant relation between birth order, the sex of siblings and the behavior in a contact sport (rugby) in a socially and culturally homogeneous setting. Second-born boys with older brothers tend to receive more yellow cards than boys with older sisters or ones that are not second-born. This result is consistent with the findings of Breining et al. (2020). Furthermore, this might be relevant for designing new policies in sports training. Coaches of child and junior teams should put more attention to the behavior of second-born players, being alert to possible  misbehaviors and helping them to learn how to control themselves.\\
A secondary result is the existence of a positive impact of weight on the number of yellow cards. This might be a consequence of, as mentioned in Section 3, the fact that heavy players usually play as forwards, who are more involved in breakdowns, where most of the offenses are committed.

\section*{Acknowledgments}
We thank the Uni\'on de Rugby del Sur for allowing us to access their records and for conducting the survey on family and education of players.

\section*{References}
\begin{itemize}
\item Bard, D.E. and Rodgers, J.L. (2003). Sibling Influence on Smoking Behavior: A Within‐Family Look at Explanations for a Birth‐Order Effect. {\it Journal of Applied Social Psychology} 33: 1773-1795. 
\item Barclay, K. and Kolk, M. (2018). Birth intervals and health in adulthood: a comparison of siblings using Swedish register data. {\it Demography} 55(3): 929--955.
\item Bautista Branz, J. (2016). Being close to Europe. Sport, social class and prestige in Argentina. {\it Revista Reflexiones} 95(1): 131--142.
\item Black, S. E., Gr\"onqvist, E. and \"Ockert, B. (2018). Born to lead? The effect of birth order on noncognitive abilities. {\it Review of Economics and Statistics} 100(2): 274-286.
\item Breining, S., Doyle, J., Figlio, D.N., Karbownik, K. and Roth, J., 2020. Birth order and delinquency: Evidence from Denmark and Florida. {\it Journal of Labor Economics} 38(1): 95-142.
\item Cameron, A.C. and P.K. Trivedi (2013). {\bf Regression analysis of count data}. Cambridge University Press.
\item Damian, R. and  Roberts, B. W. (2015). The associations of birth order with personality and intelligence in a representative sample of US high school students. {\it Journal of Research in Personality} 58: 96-105.
\item Dunn, J. and Plomin, R. (1990). {\bf Separate lives: Why siblings are so different}. Basic Books.
\item Ernst, C. and Angst, J. (1983). {\bf Birth order: Its influence on personality}. Springer-Verlag.
\item Esposito, L., Kumar, S. M. and Villase\~nor, A. (2020). The importance of being earliest: birth order and educational outcomes along the socioeconomic ladder in Mexico. {\it Journal of Population Economics}, {\tt DOI: 10.1007/s00148-019-00764-3}.
\item Forer, L. K. (1977). {\bf The birth order factor}. Pocket Books.
\item Ginja, R., Jans, J. and Karimi, A. (2020). Parental leave benefits, household labor supply, and children’s long-run outcomes. {\it Journal of Labor Economics} 38(1): 261-320.
\item Hilbe, J.M. (2011). {\bf Negative Binomial Regression}. Cambridge University Press.
\item Meller, F. O., Loret de Mola, C., Assunç\~ao, M. C. F., Schäfer, A. A., Dahly, D. L. and Barros, F. C. (2018). Birth order and number of siblings and their association with overweight and obesity: a systematic review and meta-analysis. {\it Nutrition Reviews} 76(2): 117-124.
\item Orme, J. G., and Combs-Orme, T. (2009). {\bf Multiple regression with discrete dependent variables}. Oxford University Press.
\item Romand, P. and Pantal\'eon, N. (2007). A qualitative study of rugby coaches’ opinions about the display of moral character. {\it The Sport Psychologist} 21(1), 58-77.
\item Simon, H. A. (2000). Bounded rationality in social science: Today and tomorrow. {\it Mind and Society} 1(1): 25-39.
\item Sulloway, F. J. (1996). {\bf Born to rebel: Birth order, family dynamics, and creative lives}. Pantheon Books.
\end{itemize}
\newpage
\section*{Appendix 1: Survey on family structure of rugby players.}

\begin{itemize}
	\item First and Last Name
	\item Age
	\item Club
	\item How many years have you been playing rugby? (if you started this year write 0)
	\item Educational Attainment
	\begin{itemize}
		\item Incomplete Secondary Studies
		\item Complete Secondary Studies
		\item Incomplete College Studies
		\item Complete College studies
	\end{itemize}
	\item How many siblings do you have?
	\item Enumerate your and your siblings order of birth, indicating gender, from the oldest to the youngest? (In the positions corresponding to you, write {\bf Me}).  For example: 1.Brother 2. {\bf Me} 3. Sister
\end{itemize}

\section*{Appendix 2: Regression Results}

\begin{landscape}
	\begin{table}[!htbp] \centering 
		\caption{Regression Results} 
		\label{tab:results} 
		
		\begin{tabular}{l c c c c c c c c} \\[-1.8ex]\hline 
		\hline \\[-1.8ex] 
		Variable & Standard & Robust & Robust & Robust & Robust & Robust & Cluster & Cluster\\
		\hline
		{\em Code 1} & $0.3689$ & $0.3721$ & $0.3589$ & $0.2686$ & $0.2318$ & $0.3451$ & $0.3689$ & $0.3721$ \\
		& $(0.1346)^{***}$ & $(0.1349)^{***}$ & $(0.1449)^{**}$ & $(0.1208)^{**}$ & $(0.1271)^{**}$ & $(0.1354)^{***}$ & $(0.2199)^{***}$ & $(0.2242)^{*}$ \\
		{\em Code 2} & $-0.2768$ & $-0.2603$ & $-0.3204$ & $-0.2999$ & $-0.1797$ & $-0.2261$ & $-0.2768$ & $-0.2603$ \\
		& $(0.1636)^{*}$ & $(0.1635)^{*}$ & $(0.1741)^{*}$ & $(0.2167)$ & $(0.2288)$ & $(0.1631)$ & $(0.1311)^{***}$ & $(0.1256)^{***}$ \\
		{\em Age} & $$ & $0.0082$ & $$ & $0.0023$ & $-0.0142$ & $$ & $$ & $0.0082$ \\
		& $$ & $(0.0088)$ & $$ & $(0.0113)$ & $(0.0156)$ & $$ & $$ & $(0.0067)$ \\
		{\em Inc Second} & $$ & $0.0363$ & $$ & $$ & $0.5010$ & $$ & $$ & $0.0363$ \\
		& $$ & $(0.1564)$ & $$ & $$ & $(0.2554)^{*}$ & $$ & $$ & $(0.1995)$ \\
		{\em Comp Tert} & $$ & $-0.0211$ & $$ & $$ & $0.0384$ & $$ & $$ & $-0.0211$ \\
		& $$ & $(0.1507)$ & $$ & $$ & $(0.2042)$ & $$ & $$ & $(0.1143)$ \\
		{\em Inc Tert} & $$ & $0.1947$ & $$ & $$ & $0.3207$ & $$ & $$ & $0.1947$ \\
		& $$ & $(0.1372)$ & $$ & $$ & $(0.2830)$ & $$ & $$ & $(0.2049)$ \\
		{\em Weight} & $$ & $$ & $0.0113$ & $0.0125$ & $0.0095$ & $$ & $$ & $$ \\ 
		& $$ & $$ & $(0.0055)^{**}$ & $(0.0046)^{**}$ & $(0.0079)$ & $$ & $$ & $$ \\
		{\em Height} & $$ & $$ & $-0.3169$ & $0.1856$ & $0.5356$ & $$ & $$ & $$ \\
		& $$ & $$ & $(1.1010)$ & $(1.1740)$ & $(1.4222)$ & $$ & $$ & $$ \\
		{\em Forward} & $$ & $$ & $0.1054$ & $$ & $$ & $$ & $$ & $$ \\
		& $$ & $$ & $(0.1922)$ & $$ & $$ & $$ & $$ & $$ \\
		{\em Back} & $$ & $$ & $-0.0757$ & $$ & $$ & $$ & $$ & $$ \\
		& $$ & $$ & $(0.1696)$ & $$ & $$ & $$ & $$ & $$ \\
		{\em Wing} & $$ & $$ & $$ & $0.0969$ & $$ & $$ & $$ & $$ \\
		& $$ & $$ & $$ & $(0.2201)$ & $$ & $$ & $$ & $$ \\
		\hline
		{\em Category} & NO & NO & NO & NO & YES & NO & NO & YES\\
		{\em Position} & NO & NO & NO & NO & YES & NO & NO & NO\\
		{\em Club} & NO & NO & NO & NO & NO & YES & NO & NO\\
		{\em Constant} & $0.6310$ & $0.3697$ & $0.3885$ & $-0.7304$ & $-1.3998$ & $0.8889$ & $0.6310$ & $0.3697$ \\
		& $(0.0616)^{***}$ & $(0.2315)$ & $(1.7550)$ & $(1.9911)$ & $(2.4521)$ & $( 0.1597)$ & $(0.1110)^{***}$ & $(0.1809)$ \\
		Obs & $415$ & $415$ & $310$ & $310$ & $310$ & $415$ & $415$ & $415$ \\
		\hline
		\multicolumn{6}{l}{\textit{Note:} $^{*}$p$<$0.1; $^{**}$p$<$0.05; $^{***}$p$<$0.01} & & & \\ 
	\end{tabular}	
	\end{table}  
\end{landscape}

\begin{landscape}
	\begin{table}[!htbp] \centering 
		\caption{Regression Results (CONT.)} 
		\label{tab:results2} 
		
		\begin{tabular}{l c c c c c c c c} \\[-1.8ex]\hline 
		\hline \\[-1.8ex] 
		Variable & Poisson & Poisson & Poisson & Poisson & Poisson & NB2 & NB2 & NB2\\
		\hline
		{\em Code 1} & $0.4603$ & $0.3908$ & $0.3857$ & $0.3795$ & $0.4006$ & $0.4603$ & $0.4272$ & $0.3884$ \\
		& $(0.1657)^{***}$ & $(0.1540)^{***}$ & ${0.1542}^{**}$ & ${0.1541}^{**}$ & $(0.1541)^{***}$ & $(0.1790)^{**}$ & $(0.1925)^{**}$ & $(0.1931)^{**}$ \\
		{\em Code 2} & $-0.5775$ & $-0.5061$ & $-0.4972$ & $-0.4988$ & $-0.4833$ & $-0.5775$ & $-0.5171$ & $-0.5138$ \\
		& $(0.3208)^{*}$ & $(0.2698)^{*}$ & ${(0.2701)}^{*}$ & ${(0.2703)}^{*}$ & $(0.2707)^{*}$ & $(0.2904)^{**}$ & $(0.3044)^{*}$ & $(0.3054)^{*}$ \\
		{\em Age} &  &  &  &  & $-0.0055$ &  & $-0.0048$ & $-0.01423$ \\
		&  & &  & & $(0.0128)$ &  & $(0.0151)$ & $(0.0132)$ \\
		{\em Weight} &  & $0.0142$ & $0.0138$ & $0.009$ & $0.0129$ &  & $0.0149$ & $0.0153$ \\ 
		&  & ${(0.0038)}^{***}$ & ${(0.003)}^{***}$ & ${(0.004)}^{*}$ & $(0.004)^{***}$ &  & $(0.0047)^{***}$ & $(0.0047)^{***}$ \\
		{\em Height} &  &  & $0.2594$ & $0.041$ & $0.3628$ &  &  &  \\
		&  &  & $(1.0772)$ & $(1.159)$ & $(1.1008)$ &  &  &  \\
		{\em Forward} &  &  & $0.005$ &  &  &  &  &  \\
		&  &  & $(0.1837)$ &  &  &  &  &  \\
		{\em Back} &  &  & $-0.2538$ &  &  &  &  &  \\
		&  &  & $(0.1874)$ &  &  &  &  &  \\
		$\alpha$ &  &  &  &  &  & $0.8277$ & $0.5253$ & $0.5281$ \\
		&  &  &  &  &  & $(0.1962)^{***}$ & $(0.1600)^{***}$ & $(0.1598)^{***}$ \\
		\hline
		{\em Category} & NO & NO & NO & NO & YES & NO & YES & NO\\
		{\em Studies} & NO & NO & YES & NO & NO & NO & YES & NO\\
		{\em Club} & NO & NO & NO & YES & NO & NO & NO & YES\\
		{\em Constant} & $-0.4603$ & $-1.9536$ & $-1.922$ & $-1.1922$ & $-1.8157$ & $-0.4603$ & $-1.3089$	& $-1.8866$ \\
		& $(0.0979)^{***}$ & $(1.8094)$ & $(1.8057)$ & $(1.9644)$ & $(1.9211)$ & $(0.0912)$ & $(0.6851)$ & $(0.5013)$ \\
		Obs & $415$ & $310$ & $310$  & $310$ & $310$ & $415$ & $310$ & $310$\\
		\hline
		\multicolumn{6}{l}{\textit{Note:} $^{*}$p$<$0.1; $^{**}$p$<$0.05; $^{***}$p$<$0.01} & & & \\ 
\end{tabular}	
	\end{table}  
\end{landscape}

\end{document}